\documentstyle[general_cite,psfig]{mn}    

\title[The L--T relation of high-z clusters]{The WARPS survey -- IV: The X-ray luminosity--temperature relation of
  high redshift galaxy clusters}

\author[B. W. Fairley et al.]
{B. W. Fairley$^{1,8}$, L. R. Jones$^1$, C. Scharf$^{2}$, H. Ebeling$^{3}$,
  E. Perlman$^{2}$, \cr D. Horner$^{4,7}$, G. Wegner$^{5}$ and M. Malkan$^{6}$ \\
$^1$ School of Physics and Astronomy, University of
Birmingham, Birmingham, B15 2TT, UK\\
$^2$ Space Telescope Science Institute, Baltimore, MD 21218,
USA\\
$^3$ Institute for Astronomy, 2680 Woodlawn Dr, Honolulu, HI
96822, USA\\
$^4$ Laboratory for High Energy Astrophysics, Code 660,
NASA/GSFC, Greenbelt, MD 20771, USA\\
$^5$ Dept. of Physics \& Astronomy, Dartmouth College, 6127
Wilder Lab., Hanover, NH 03755, USA\\
$^6$ Dept. of Astronomy, UCLA, Los Angeles, CA 90024, USA\\
$^7$ University of Maryland, College Park, MD 20742-2421, USA\\
$^8$ Email: bwf@star.sr.bham.ac.uk\\}
\date{Accepted .....................; Received .....................;
in original form .......................}

\begin{document}                       

\maketitle

\begin{abstract} 

We present a measurement of the cluster X-ray luminosity-temperature
relation out to high redshift ($z \sim 0.8$). Combined $ROSAT$ PSPC spectra of 91
galaxy clusters detected in the Wide Angle $ROSAT$ Pointed Survey (WARPS)
are simultaneously fit in
redshift and luminosity bins. The resulting temperature and
luminosity measurements of these bins, which occupy a region of the high redshift L-T
relation not previously sampled, are compared to existing measurements at
low redshift in order to
constrain the evolution of the L-T relation. We find
a best fit to low redshift ($z<0.2$) cluster data, at $\rm T > 1 \, keV$, to be
$\rm L \propto T^{3.15\pm0.06}$. Our data are consistent with no evolution
in the normalisation of the L-T relation up to $z \sim 0.8$. Combining our
results with $ASCA$ measurements taken from the literature, we find
$\eta=0.19\pm0.38$ (for $\Omega_0=1$, with $1\sigma$ errors) where $\rm
L_{Bol} \propto {(1+\it z)}^{\eta} \rm T^{3.15}$, or $\eta=0.60\pm0.38$ for
$\Omega_0=0.3$. This lack of evolution is
considered in terms of the entropy-driven evolution of clusters. Further
implications for cosmological constraints are also discussed.

\end{abstract}

\begin{keywords}
galaxies: clusters: general -- methods: data analysis -- surveys --
  X-rays: galaxies -- cosmology: observations
\end{keywords}

\section{Introduction}

Clusters of galaxies, as the largest virialised objects in the
Universe, allow us a unique insight into the formation and evolution of
mass clustering on cosmological time-scales. The X-ray emission from galaxy
clusters originates from hot ($\rm T \sim 10^{7-8}$ K) intra-cluster gas
which, in relaxed systems, is in hydrostatic equilibrium with the cluster's
total gravitational field. Since the temperature of the
gas is proportional to its mass, the observed correlation between X-ray
luminosities and temperatures implies a relationship between
cluster baryon mass and total mass. Thus, studying the evolution of the L--T
relation probes the interrelated evolution of these mass components.
One of the first attempts at modelling these properties assumed self-similar
evolution of both the gravitational potential and the baryonic intracluster
medium (ICM) at varying cosmological epochs (\pcite{kai86}).

These self-similar predictions are in conflict with observational
evidence. The strong positive evolution of the
cluster X-ray luminosity function (XLF) originally predicted (assuming
realistic cosmological parameters) is in stark
contrast to the negative evolution reported by \scite{gio90} and
\scite{hen92} for X-ray luminous clusters. The prediction is also in
conflict with the results of more recent cluster
surveys which find no significant evolution, at
any luminosity, out to moderate redshift ($z \sim 0.3$, \pcite{ebe97};
\pcite{deg99}), and again no significant (or mild negative) evolution at
low and intermediate luminosities and high redshifts (\pcite{ros95};
\pcite{bur97}; \pcite{vik98a}; \pcite{jon98b}). To account for such
discrepancies, modifications were suggested to the
self-similar theory. \scite{kai91} and
\scite{evr91} modelled a 'pre-heated ICM', in which an initial injection of
energy broke the direct self-similar scaling by introducing an entropy floor
in cluster cores. It also brought theoretical predictions of cluster properties into
better agreement with observational constraints.

Tests of gravitational self-similar scaling are thus
fundamental to the understanding of cluster evolution. One such test used
in this analysis is the study of the cluster X-ray luminosity--temperature
(L--T) relation, which was predicted by the original self-similar model to be
described by an $\rm L \propto T^2$ law. However, a relationship
closer to $\rm L \propto T^3$, has been found by several cluster surveys and
compilations (e.g. \pcite{edg91}; \pcite{dav93}; \pcite{whi97}), whilst
even further steepening of
the relation is required to agree with galaxy group measurements
(\pcite{pon96}; \pcite{hel00}). In addition to predicting the slope of the L--T relation, self-similar
theory also predicts significant evolution in the normalisation of this
relation at different redshifts. However, the evolution predicted is
reduced in low density cosmologies, or by significant entropy injection at
early epochs (e.g. \pcite{kai91}; \pcite{cen94}; \pcite{kay99}). 

Attempts at constraining evolution in the L--T relation, at redshifts above
$z \sim 0.1$, are complicated by the inherent difficulty of temperature
estimation from often poor signal-to-noise data. \scite{mus97}, using $ASCA$
data from a sample of 38 clusters at $z > 0.14$, found no significant
evolution out to $z \sim 0.3$. Prior to this work, the only attempt
at quantifying evolution of the L--T relation to $z \sim 0.3$ was that
of \scite{hen94}, whose results were also
consistent with no evolution. \scite{hen94} used low signal-to-noise
spectra of 67 clusters observed with the $Einstein$ observatory,
each individually spectrally fitted, and then averaged together in redshift
bins. Recently \scite{don99} have used $ASCA$ measurements of clusters drawn from the
$Einstein$ Medium Sensitivity Survey (EMSS, \pcite{gio90}; \pcite{hen92}) to extend the conclusion of no
evolution to $z > 0.5$.

In this paper we present an L--T relation based on $ROSAT$ spectra of WARPS
clusters, in various redshift and
luminosity bins. Although this procedure is analogous to the one adopted by
\scite{hen94}, a more sophisticated simultaneous fitting method
is adopted here. We discuss the data
analysis techniques used to obtain temperature estimates from often low
signal-to-noise $ROSAT$ PSPC data, and present results from simulated
spectra to demonstrate their validity. In order to allow self-consistent
discussion of any
evolutionary effects in our relation, we compare our results
with a local L--T relation over a wide range of temperatures and
luminosities. 
 
\section{Sample selection}

Traditional studies of galaxy clusters have relied
on optically selected cluster
samples. Whilst these surveys were often large, they suffered from biasing
caused by projection effects and erroneous identifications (\pcite{luc83};
\pcite{fre90}; \pcite{str91}). The recent trend, therefore, has been to
base large cluster surveys on X-ray
selection techniques. This takes advantage of the X-ray emission from
diffuse gas, trapped in a cluster's gravitational potential well,
representing direct evidence of a three-dimensionally bound system. Several
recent serendipitous cluster surveys have been compiled, all based
on archival $ROSAT$ PSPC data (e.g. \pcite{cas95}; \pcite{ros95};
\pcite{col97}; \pcite{vik98b}; \pcite{rom00}).

The Wide Angle $ROSAT$ Pointed Survey (WARPS, \pcite{sch97}) is also based
on archival $ROSAT$ PSPC data. It is a serendipitous cluster survey which
uses a sophisticated source detection algorithm, {\small VTP} \normalsize (\pcite{ebe93}), capable of detecting arbitrarily shaped emission, down to
very low surface brightness levels. An extensive optical follow-up program then
spectroscopically confirms the concordance of galactic redshifts within
each candidate cluster (\pcite{jon98a}). When complete, the WARPS survey
will have generated a statistically complete, flux-limited catalogue
covering a solid angle of more than 73 $\rm deg^2$ and including
approximately 150 X-ray selected galaxy clusters.

The sample used in this paper comprises 91 clusters identified in the WARPS
survey, with redshifts in the range from $z=0.11$ to $z=0.833$. The WARPS
survey is flux-limited at $\rm 6 \times
10^{-14} \: erg \: s^{-1} \: cm^{-2}$ (0.5--2 keV) in total cluster flux and
is restricted to $ROSAT$ pointings
of more than 8-ks exposure at $\rm |b|>20\deg$. Only the most sensitive area
of the PSPC detector, at off-axis angles of $3<r<15$ arcmin, is used.
In addition to the archival
$ROSAT$ PSPC data, deep optical imaging of the clusters, optical spectroscopy of
all the brightest cluster galaxies (BCGs) and in some
cases $ROSAT$ HRI data are available (\pcite{jon98a}; \pcite{per00}).

\section{Data reduction}

\subsection{Spectral fitting}

For this work, the archival $ROSAT$ PSPC data, containing the 91 WARPS
clusters selected, were initially reduced and analysed using the
Starlink {\small ASTERIX} \normalsize package (\pcite{all95}). The data were screened to minimize particle
background rates by restraining the Master Veto Rate housekeeping parameter to a
range of 0 to 170. The Aspect Error and Accepted Event Rate parameters were
also limited to ranges of 0 to 2 and 0 to 30 respectively. The observation
time thus removed was on average 5 per cent. The resulting cleaned data were then used to
produce binned spectral cubes, with one spectral and two spatial
dimensions, from $ROSAT$ channels 11 through 231, with
energy ranges of 0.114 to 2.295 keV

Background subtraction for each field was achieved by creating a model background spectral
cube. Firstly an annulus was selected between 9 and
15 arcmin off-axis. This area is included in the WARPS analysis and was
selected to minimize vignetting extrapolation errors in the background
model. The background cube was projected to an image and
searched for point sources using the {\small ASTERIX PSS} \normalsize algorithm. Removal of the
sources resulted in a new background model. This process was iteratively repeated until no more point
sources were detected. The point source free cube was then
further edited manually to eliminate any extended sources, such as cluster
emission. The background level of this cube was then extrapolated over all
off-axis angles, using the known energy-dependent vignetting behaviour of the detector, to
produce a final background model.

After background subtraction, a spectrum of the cluster was obtained by
projecting a selected area of the spectral cube on to the energy axis, before correcting
for energy-dependent vignetting and exposure effects. The circular
extraction regions around the clusters had initial radii of 700 and 900
kpc, for $z<0.5$ and $z>0.5$ respectively (we assumed $\rm H_0=50 \:
km \: s^{-1} \: Mpc^{-1}$ and $q_0=0$). These radii were manually
re-selected, where required, to eliminate contamination from nearby
non-cluster sources.
A background spectrum from the same area of sky, obtained from the
source-free background model, was then associated with the cluster
spectrum to allow spectral fitting using the maximum-likelihood C-statistic
(\pcite{cas79}). The number of counts in the cluster spectra ranged from 25
to 1001 (0.11--2.3 keV) with a median value of $\sim 120$ counts. The
signal-to-noise ratios of the data varied dramatically, depending
primarily on background level, exposure time and extraction
radius. The average signal-to-noise level was $\sim 6.8$ in the
0.1--2.3 keV band, but was significantly higher in the 0.5--2 keV band used
in the initial detection (\pcite{sch97}). Thus C-statistic
fitting was selected to avoid biases inherent in $\chi^2$ fitting for low
count-per-bin data (\pcite{nou89}). 

The spectral models fitted to the data were the  \scite{ray77} and {\small MEKAL} \normalsize (\pcite{mew86};
\pcite{kaa92}) codes for an optically thin thermal plasma. The column density 
was fixed at the Galactic value determined from radio measurements (\pcite{sta92}), and a
metallicity of 0.3 $\, Z_{\odot}$ was assumed. It was found that both codes gave very
good agreement; the results presented here are from the {\small MEKAL} \normalsize model.

Due to the very low number of counts in many of the
individual cluster spectra, the spectral fits were often poorly
constrained, with several not converging. In addition fits to a number of
high redshift, low signal-to-noise spectra converged to low temperatures with
small errors (e.g. T=$1.3 \pm 0.1$ keV). This was of concern as, according to
the observed, low redshift, L--T relation, this implied luminosities too low
for these clusters to be detected above the flux-limit of the survey. 

There are two primary characteristics of the data which dominate the
accuracy and reliability of the cluster
temperature measurements, and cause the effect discussed in the previous
paragraph. Firstly, the signal-to-noise of the cluster
above the background, and secondly, the accuracy to which the background level
can be determined. To investigate possible
systematic biases caused by poor photon statistics and erroneous background
estimates, extensive simulations were
performed, using the {\small ASTERIX SSIM} \normalsize package. Fig.~\ref{fig:fithist1} shows
spectrally fitted temperatures from 100 datasets all simulated from a 5keV
{\small MEKAL} \normalsize model. Each cluster was simulated as excess emission on top of a PSPC background of
800 counts (0.1--2.3 keV, in the extraction region).
This background is brighter than any found in the actual observations;
hence the background level can be accurately determined in these
simulations. The total number of counts is varied according to Poisson
statistics. Spectral fits were performed on the emission after the
background was subtracted. It can be seen from the histogram that for the
spectra (in
this case with an average of $\sim 190$ counts in the 0.1--2.3 keV band,
giving a signal-to-noise of 6.0), three distinct temperature populations were
found. Spectral fits that were poorly constrained had the best-fitting temperatures pegged
at a maximum value (17 keV in this case). For a different subset of spectra
the fit essentially recovered the simulated temperature. However, a third
population of low temperature fits with small errors was also found, as
observed with some of the $ROSAT$ datasets. This is a demonstration of the
bias introduced with low signal-to-noise in the cluster spectra, but with
the background level reasonably well determined.

The combination of the biases due to low signal-to-noise and poor
background statistics, is demonstrated in
Fig.~\ref{fig:fithist2}, which shows the results from 200
simulated spectra, each containing $\sim 120$ counts (0.1--2.3 keV). The
background level in this simulation was 
tailored to give a signal-to-noise values around 7 (i.e. to have $\sim175$
counts, 0.1--2.3 keV). Here the individual fits
cannot recover the simulated temperature of 5 keV.

A method of increasing the signal-to-noise of the spectra was thus required
to enable accurate temperature estimates to be made. This was achieved via
simultaneous multiple fitting of data sets in binned groups.

\subsection{Multiple fitting}

\begin{figure*}
\begin{minipage}{14cm}
\psfig{file=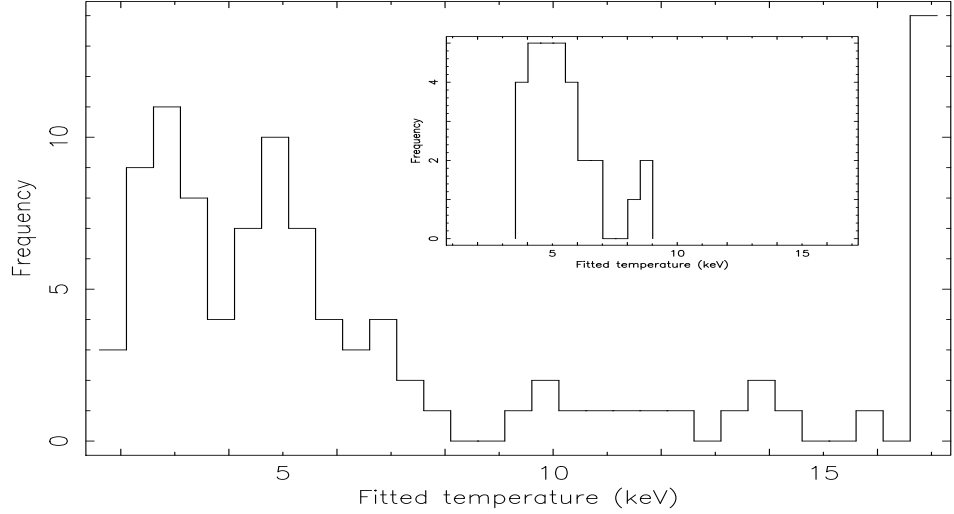,width=14cm,height=9cm,angle=270}
\caption{\label{fig:fithist1}Spectrally fitted temperature estimates for 100 simulated 5 keV
  clusters with $\sim 190$ PSPC counts (0.1--2.3 keV). Note the three
  distinct populations at low temperatures, $\sim5$ keV and pegged at 17
  keV. Inset is the recovered temperature distribution after multiple
  fitting of sets of 5 such clusters.} 
\end{minipage}
\end{figure*}

\begin{figure*}
\begin{minipage}{14cm}
\psfig{file=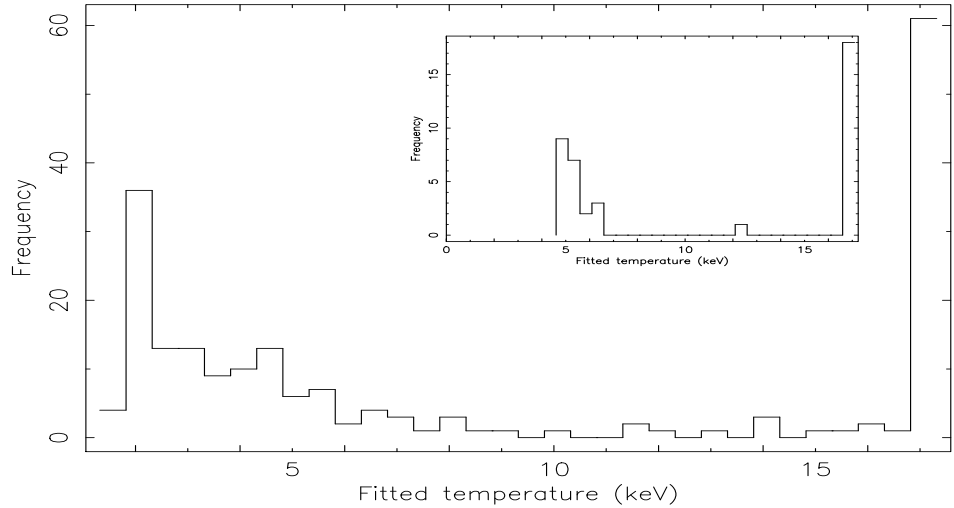,width=14cm,height=9cm,angle=270}
\caption{\label{fig:fithist2}Spectrally fitted temperature estimates for 200 simulated 5 keV
  clusters with under $\sim 120$ PSPC counts (0.1--2.3 keV). The individual
  spectral fits cannot recover the simulated temperature. Inset is the
  recovered temperature distribution after multiple fitting of sets of 5
  such clusters.} 
\end{minipage}
\end{figure*}

The effect of multiply fitting poor signal-to-noise datasets can be seen in
the inset panels in both Fig.~\ref{fig:fithist1} and
Fig.~\ref{fig:fithist2}. These show temperatures estimated from simultaneously
fitting subsets of five of the simulated datasets discussed in the previous
section. In most cases, values around the simulated temperature of 5 keV
were found by these fits.

As can be seen from Fig.~\ref{fig:fithist2}, the primary effect of
reducing the cluster count level, together with incorporating a low
statistical accuracy background measurement, is to produce a number of simultaneous
fits that are not constrained and yield the maximum allowed temperature of
17 keV. As the simulated signal-to-noise level is
similar to the poorer PSPC spectra used in this work, we might
expect to see this effect when using this method on WARPS clusters. No such
trend was observed though. This was probably because in the real
measurements a large background area was sampled in order to derive an
accurate background level. This is in contrast to the simulations, where a
much noisier background measurement resulted from using an area $\sim100$
times smaller (as required by the {\small SSIM} \normalsize program). This
difference also accounts for the increased accuracy of the
simulations with higher backgrounds, but similar signal-to-noise. In the
analysis of PSPC data, the large area used in the estimation of the individual background level
for each cluster serves to minimize the second of the two biases discussed
in the previous section.

Numerous simulations were carried out to test the effectiveness of the
method. Gaussian distributions in both
temperature (e.g. $\rm \pm1 keV$ at 90 per cent confidence) and metallicity
($\pm0.1 \, Z_{\odot}$), around the simulated values, were included.
These extended simulations also recovered a correct
temperature distribution, though with a larger spread spanning around
$\rm \pm2keV$. The temperature errors obtained from the fits, however,
generally encompassed the distribution range. 

From all the simulations, we conclude that multiple fits of five spectra,
with each individual spectra containing 100--200 counts (i.e. a
signal-to-noise ratio of 6--7 for PSPC spectra), with one free parameter,
and the background statistically well determined,
produce temperatures in which the level of any systematic bias is smaller
than the random error. There may be an additional temperature uncertainty
at high temperatures due to PSPC calibration uncertainties. \scite{mar97}
found evidence of significant differences between PSPC and $ASCA$ temperatures for
T $\sim5$ keV and higher, although the size of the differences are around the same
size as our statistical error.

Multiple fitting then, was the method adopted for analysis of the PSPC spectra. In order to
produce averaged spectra, we binned the clusters into redshift
and luminosity bins. Where possible at least five clusters were included in
each bin. Simultaneous multiple fitting was done within Version 10.0 of the {\small
  XSPEC} \normalsize fitting package (\pcite{arn96}) in
the {\small XANADU} \normalsize X-ray software suite. This enables several sets of data to be
fitted to the same model, with certain parameters either frozen or tied to
be the same for all datasets. In our case, the clusters were fitted to the
same temperature in the plasma code, but with individually varying
normalisations. Again, column densities were fixed at Galactic
values for each cluster and metallicities were fixed at 0.3 $\, Z_{\odot}$.

The main complication arose from the need to use maximum likelihood
statistics. To correctly fit using the Cash statistic, {\small XSPEC} \normalsize requires a non-background
subtracted cluster spectrum, with a pre-fitted background component. Thus
multiple fitting first involved individually fitting the relevant
background for each cluster. The model used was an absorbed combination of
a {\small MEKAL} \normalsize plasma, a power law and a Gaussian emission line profile, with an initial
energy of 0.54 keV (as used by \scite{bra94} to model the X-ray background
spectrum).

\subsection{Luminosities}

Flux estimates in the 0.5--2 keV band were obtained for each cluster from the
individual fits to the spectra. These were then k-corrected, and corrected for low surface
brightness emission outside the selected area assuming a King profile with $\beta=\frac{2}{3}$ and
cluster core radius estimated from the {\small VTP} \normalsize data, as described by \scite{sch97}. The fluxes were then
converted to luminosities, using values of $\rm H_0=50 \: km \: s^{-1} \: Mpc^{-1}$
and $\rm q_0=0$. These luminosity estimates were then used for the binning of the clusters within each
redshift range. 

Due to the need for an accurate bolometric conversion from 0.5--2 keV luminosities,
which relies on temperature estimates, the bolometric luminosities of each
cluster used in
the final results were obtained after the multiple fit. Each cluster
was bolometrically corrected using the characteristic temperature of the
bin, before again being k-corrected and convolved through a King profile
correction to produce accurate luminosities. The typical bolometric
correction resulted in a factor of 2--3 increase from the 0.5--2 keV
luminosity. The cluster luminosities within the bin were then
averaged to provide a characteristic bin luminosity.

The steep slope and large inherent scatter of the low redshift L--T relation
(e.g. \pcite{dav93}; \pcite{whi97}), heightens the importance of restricting the
range of luminosities within each luminosity bin. A large spread in
luminosities could result in large temperature differences in the clusters
contained within any bin. Which would then invalidate the simultaneous
fitting method employed. The flux-limited nature of the WARPS data meant
that a large spread in luminosities was found only when attempting to bin
together a number of the more luminous clusters within each redshift range. In
two cases, for instance, there was no other
cluster of similar luminosity to the most luminous
cluster, and so this cluster was individually fitted for
temperature. Both
WARPJ1416.4+2315 at $z=0.137$ and WARPJ1418.5+2510 at $z=0.294$ had more
than 800 counts in the 0.1--2.3 keV band, allowing a robust temperature
fit. This also had the effect of reducing the largest spread of
signal-to-noise ratios within any bin to less than a factor of two.

\section{Results}

\subsection{X-ray luminosity--temperature relations}

\begin{figure*}
\begin{minipage}[c]{14cm}
\psfig{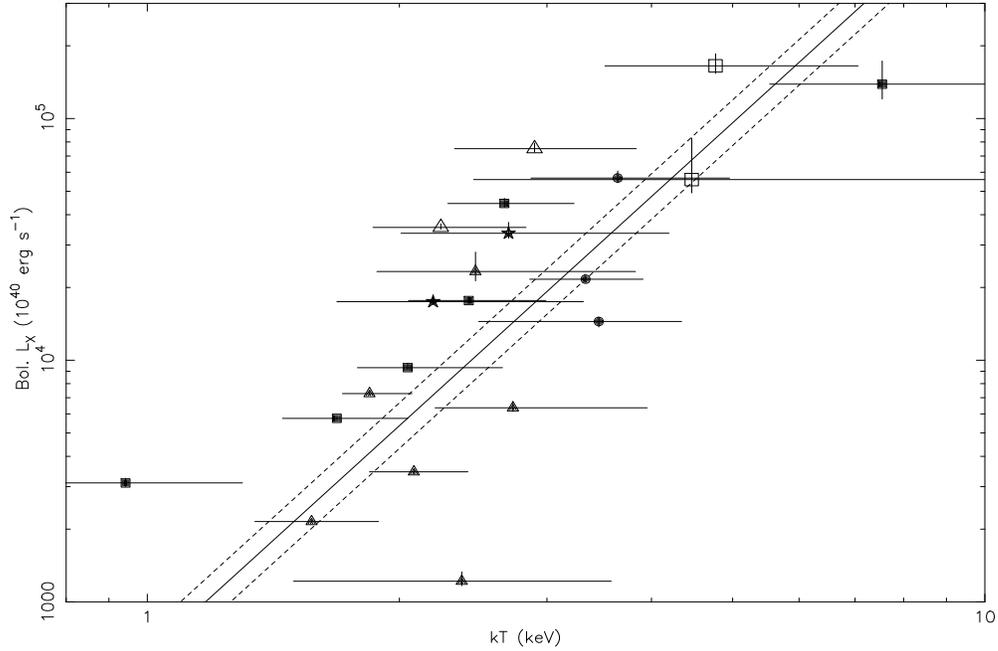}
\caption{\label{fig:lt}The L--T relation from binned PSPC spectra of WARPS
  clusters. Filled triangles represent clusters in the
  $0.1<z<0.2$ bins. Filled squares represent $0.2<z<0.3$, circles $0.3<z<0.4$, stars
  $0.4<z<0.5$, open triangles $0.5<z<0.6$ and open squares clusters of
  $z>0.6$. Each point is shown with $1\sigma$ temperature and luminosity
  errors. Also shown is the best-fitting line to the data shown in figure 4 at
  T$\: > 1$ keV, $\rm L \propto
  T^{3.15}$ with its 90 per cent confidence normalisation errors.} 
\end{minipage}
\end{figure*}

Fig.~\ref{fig:lt} shows the L--T relation from binned PSPC spectra of
WARPS clusters. Different symbols
represent the redshift bins used. The luminosity errors are negligible in most
cases, and primarily due to temperature uncertainties affecting the
bolometric correction.

\begin{figure*}
\begin{minipage}[c]{14cm}
\psfig{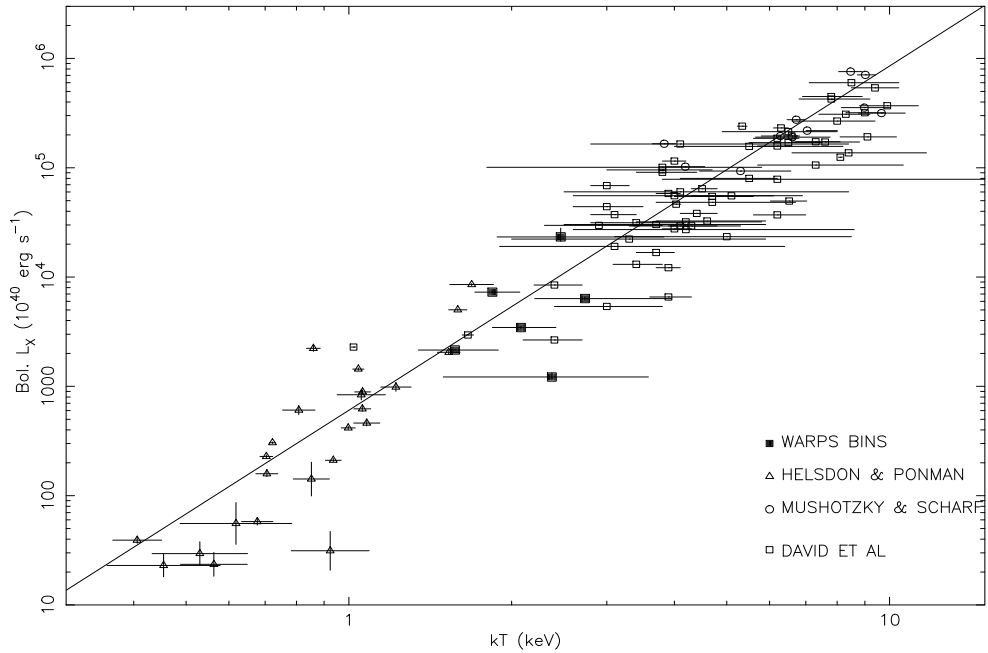}
\caption{\label{fig:lowzlt}L--T relation for $z<0.2$. Triangles represent groups from Helsdon
  \& Ponman (2000),
  filled squares are WARPS bins, open circles are clusters from Mushotzky \&
  Scharf (1997) and open squares
  are clusters from David et al. (1993). The line shown is the best-fitting line
  to clusters with $\rm T>1 keV$, $\rm L \propto T^{3.15}$.} 
\end{minipage}
\end{figure*}

\begin{figure*}
\begin{minipage}[c]{14cm}
\psfig{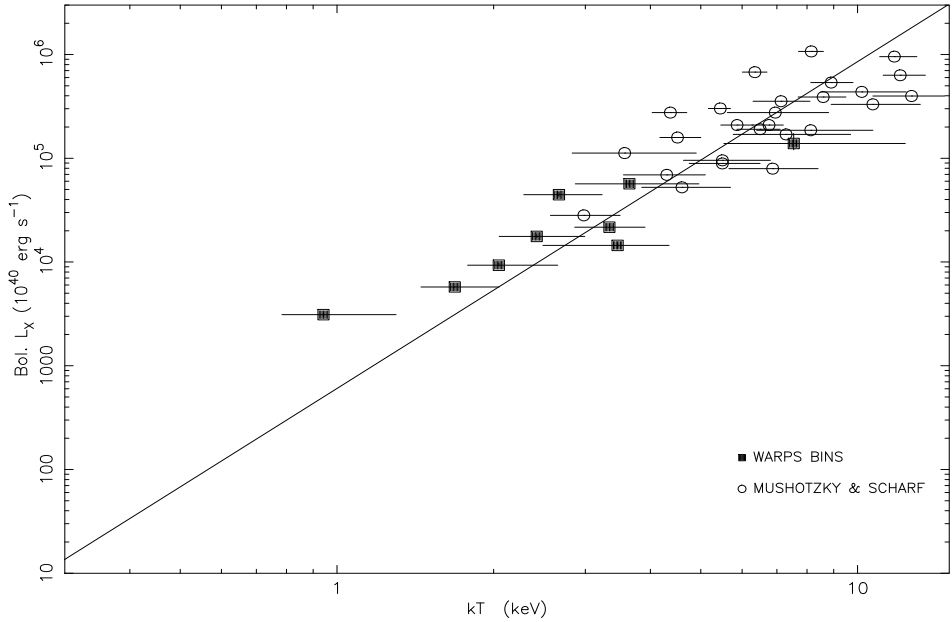}
\caption{\label{fig:24zlt}L--T relation for $0.2<z<0.4$. Filled squares represent WARPS
  bins, open  circles are clusters from Mushotzky \&
  Scharf (1997). The line shown is
  the best-fitting line from the $z<0.2$ sample.} 
\end{minipage}
\end{figure*}

\begin{figure*}
\begin{minipage}[c]{14cm}
\psfig{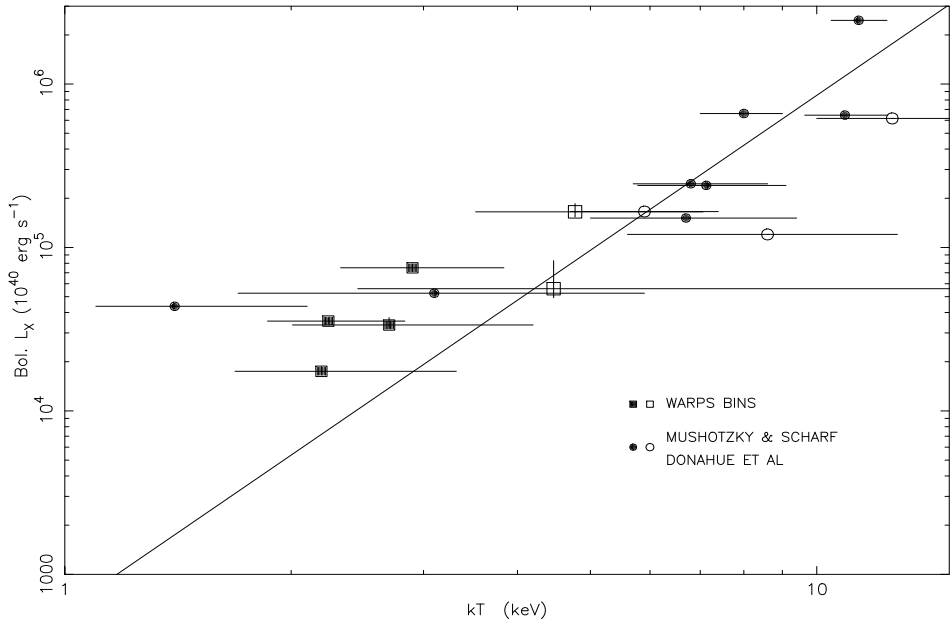}
\caption{\label{fig:46upzlt}L--T relation for $0.4<z<0.6$ and $z>0.6$. Squares represent WARPS
  bins, circles are clusters from Mushotzky \&
  Scharf (1997) and recent results from Donahue et al. (1999). In both cases
  solid points represent datapoints from the $0.4<z<0.6$ range, and open
  points $z>0.6$. The line shown is
  the best-fitting line from the $z<0.2$ sample.} 
\end{minipage}
\end{figure*}

To put this L--T relation in context, Fig.~\ref{fig:lowzlt} through 
Fig.~\ref{fig:46upzlt} show the
WARPS data points superposed on cluster L--T measurements drawn from the
literature. Data are included from a number of different
sources, including \scite{dav93}, \scite{mus97} and recent
results from \scite{don99}.  The \scite{don99} dataset includes cluster measurements from a
number of other works (\pcite{don96}; \pcite{hen97}; \pcite{hat97}; \pcite{don98}).
Data from a recent survey of 24 galaxy groups showing diffuse X-ray
emission (\pcite{hel00}), are also included.

To enable us to discuss any possible evolution within the WARPS L--T relation,
it is important to have a suitable low redshift relation to compare to.
\scite{mus97} have found no evolution in the L--T relation out to $z
\sim 0.3$. So we include all clusters at $z<0.2$ in our low
redshift sample (Fig.~\ref{fig:lowzlt}). The WARPS points are in good agreement
with previous measurements and help fill the gap in the L--T observations at
T=1--3 keV.

In order to define a fiducial low-redshift L--T relation, we obtained the
best-fitting power law to the data in Fig.~\ref{fig:lowzlt} at $\rm T>1 \:
keV$. Below T=1 keV, there may be a systematic trend towards lower
luminosities, away from the best-fitting line obtained at higher temperatures
(\pcite{pon96}; \pcite{hel00}). To allow for the systematic error
intrinsic to the use of datasets from three different X-ray observatories
and measured over different energy ranges, a minimum error of 5 per cent was
assigned to each cluster temperature where the existing
error was not already greater than this value. The best fitting slope was
then found using $\chi^2$ minimisation on the temperature axis.
For $\rm L_{Bol}=CT^{\alpha}$, we find $\alpha = 3.15 \pm 0.06$ and $\rm C = 6.04
\pm 1.47 \times 10^{42} erg s^{-1}$ (90 per cent confidence regions). Most
previous cluster L--T relations had a best-fitting slope of around $\rm L
\propto T^{2.8 \pm 0.2}$ (e.g. \pcite{mus97}, \pcite{dav93}). Our result
has a slightly steeper
slope primarily due to the inclusion of less luminous clusters not
analysed in previous investigations.

\subsection{Measuring evolution}

For the purposes of constraining evolution in the normalisation of the
WARPS L--T relation we assume a fixed slope of $\alpha = 3.15$ (section
4.1). We then further parametrize the L--T relation as $\rm L_{Bol} = C
{(1+\it z)}^{\eta} \rm T^{\alpha}$ where $\rm C = 6.04 \times 10^{42}
erg s^{-1}$, the best-fitting normalisation of the local L--T relation. The $\eta$
term then parametrizes evolution at any redshift $z$.

We now consider the WARPS measurements in the redshift ranges shown in
Fig.~\ref{fig:24zlt} and Fig.~\ref{fig:46upzlt}. Assuming the best-fitting
slope from the low redshift clusters, we again fit for normalisation by
minimising the $\chi^2$-statistic. Again only measurements above 1 keV were
included in the fit. Taking the central redshift of these bins
allows us to evaluate $\eta$ and its associated error. Most theoretical
studies discuss evolution in the L--T relation
in terms of an $\Omega_0=1$ cosmology. Table~\ref{tab:etaclose} displays our
evolutionary constraints ($\eta$ values) found for the three $z>0.2$
redshift ranges assuming $\Omega_0=1$. Wherever a different cosmology is
specified, we corrected cluster luminosities accordingly (we assume $\rm H_0=50 \: km
\: s^{-1} \: Mpc^{-1}$ throughout).

In order to further constrain evolution in the L--T plot, the number of data
points in each redshift bin must be increased. Using datasets from
\scite{mus97} and Donahue et al. (1999) in conjunction with the WARPS bins, we
were able to place tighter constraints on evolution in the L--T
plot. The data used are represented in Fig.s~\ref{fig:24zlt} and
\ref{fig:46upzlt}, and the resulting $\eta$ values are listed in Table~\ref{tab:etaclose}.

\begin{table}
\begin{minipage}[c]{8cm}
\center{\caption{\label{tab:etaclose}$\eta$ values for WARPS bins and WARPS bins combined with Mushotzky \&
  Scharf (1997) and Donahue at al. (1999) measurements, assuming an $\Omega_0=1$ cosmology. Errors shown are
  $1\sigma$. A weighted average is also given.}} 
\begin{tabular}{|c|c|c|} \hline
$z$ range & WARPS $\eta$ values & All data $\eta$ values \\ \hline
0.2--0.4 & $0.82\pm1.01$ & $0.87\pm0.55$ \\
0.4--0.6 & $2.64\pm1.12$ & $0.56\pm0.52$ \\
0.6+ & $0.45\pm1.99$ & $-1.46\pm0.86$ \\ \hline
Average= & $1.42\pm0.83$ & $0.19\pm0.38$ \\ \hline
\end{tabular}
\end{minipage}
\end{table}

It can be seen from Table~\ref{tab:etaclose} that in general the WARPS
measurements are consistent with no evolution of the normalisation of the
L--T relation (though one redshift range provides marginal evidence for
evolution). When the
WARPS points are combined with other existing cluster measurements the
results are also consistent with no evolution. To quantify the level of
evolution measured we give the averaged $\eta$ value of the three separate
redshift range measurements. The averages have been
weighted by the inverse errors. The
averaged $\eta$ value of the combined dataset for $\Omega_0=1$
($\eta=0.19\pm0.38$), is consistent with no evolution to $z
\sim 0.8$. Changing the cosmology used in the fit to an open
$\Omega_0=0.3$, $\lambda_0=0$ Universe modifies the
conclusion only slightly, producing $\eta=0.60\pm0.38$, still consistent with no
evolution at around the $1.5 \sigma$ level.

\section{Discussion}

\subsection{Theoretical relations}

The implications of the evolution parameter, $\eta$, can now be discussed
in terms of evolution of X-ray temperatures and luminosities in the L--T
relation. Here we focus primarily on comparisons with the entropy driven
evolution model of \scite{bow97} and its extension to various
cosmologies (\pcite{kay99}). In this model the evolution of the
X-ray properties of clusters is split into contributions from the
gravitational heating of the ICM and from purely gas related
processes, such as shock heating or radiative cooling.
Assuming a minimum entropy (or entropy floor) in cluster cores allows
deviations from the pure
self-similar scaling of the cluster ICM with the underlying dark matter
distribution (e.g. \pcite{kai91}; \pcite{evr91}). The entropy driven model
further allows the exploration of the regime between the purely self-similar
scaling predictions of \scite{kai86} and the effect of a pre-heated ICM.

As discussed by \scite{bow97}, the heating and cooling of the ICM gas can be
described in terms of evolution of the core entropy, if we assume that the
slope of the primordial density
fluctuation power spectrum, $n$, controls the gravitational growth of
clusters. Parameterizing this evolution in the
form
$s_{\rm min} \, = \, s_{\rm min}(\it z=\rm 0) + \it c_v \, \epsilon \, \rm ln(1+\it z)$
describes the level of the entropy floor, $s_{\rm min}$,  at any redshift,
$z$. The term $\epsilon$ thus describes how this entropy evolves ($c_v$ is
just the specific heat capacity of the gas at constant volume). A negative value
represents continual shock heating of the ICM during cluster collapse and $\epsilon>0$ describes a
cluster core dominated by radiative cooling, whilst $\epsilon=0$ is
the constant entropy model (e.g. \pcite{evr91}). It is then possible to
derive a functional form for the L--T relation, with evolution of its
normalisation parametrized in terms of $n$, and $\epsilon$. If we assume
the density profile parameter is equal to the canonical value of,
$\beta=2/3$, then both the \scite{bow97} and \scite{kay99} scaling relation
models give an L--T relation of the form $\rm L_{Bol} \propto
{(1+\it z)}^{-3\epsilon/4} T^{2.75}$. Although we find a slightly different
value of the slope of the L--T relation than this, due to the assumption of
weak self-similarity (section 2.4, \scite{bow97}), we are only concerned
with the evolution of normalisation of the relation. We therefore use a
fixed slope of $\rm L_{Bol} \propto T^{3.15}$ when dealing with these
scaling relations. The main effect of the difference in slope is to
introduce a slight degeneracy between $n$ and $\epsilon$ when calculating the
expected normalisation of the L--T relation (e.g. fig. 2, \pcite{bow97}).

\begin{figure*}
\begin{minipage}[c]{14cm}
\psfig{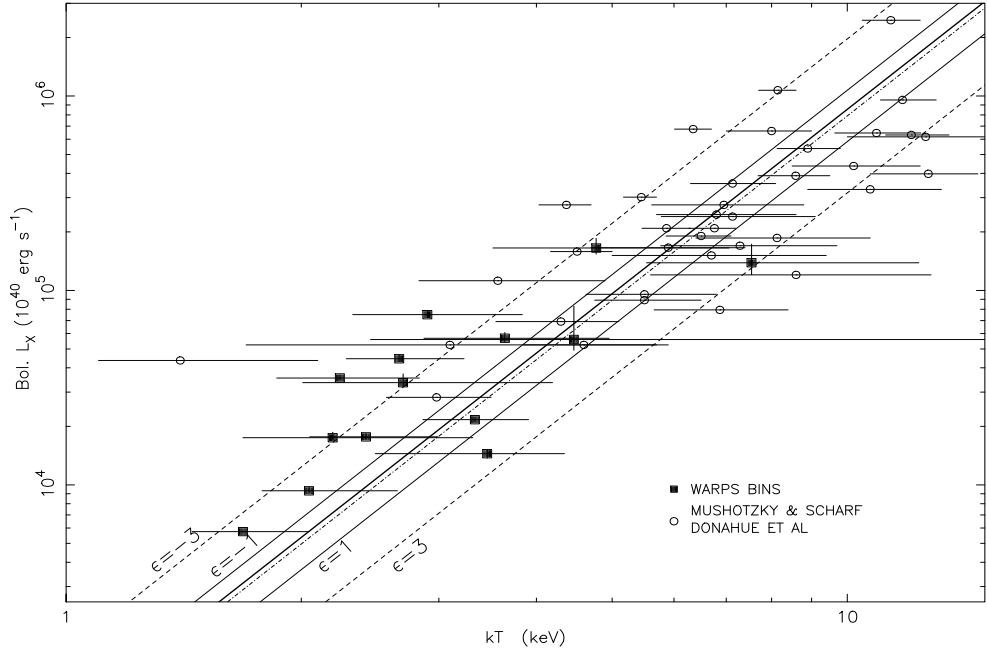}
\caption{\label{fig:kayeps}The L--T relation data above $z=0.2$, with normalisation predictions
  from Kay \& Bower (1999). WARPS points are shown as
  solid squares and the combined data of Mushotzky \& Scharf (1997) and
  Donahue et al. (1999) are displayed as open circles. The thick line represents
  the low redshift best-fitting relation. Plotted on the graph are predictions of
  the normalisation of the L--T relation assuming $\Omega_0=0.3$ and
  $n=-1.6$. Lines are calculated for a redshift of $z=0.5$, assuming 5
  different values of $\epsilon$. The dot-dashed line represents
  $\epsilon=0$, the thin solid lines $\epsilon=1$ and $\epsilon=-1$, and
  the dashed lines $\epsilon=3$ and $\epsilon=-3$. For illustrative
  purposes the lines have been converted back to a $\rm q_0=0$ cosmology,
  to match the datapoints displayed in Fig. 3 through Fig. 6.} 
\end{minipage}
\end{figure*}

We can use the scaling
relations to predict the evolution in normalisation of the L--T relation in
terms of $n$ and $\epsilon$. An example is shown in
Fig.~\ref{fig:kayeps}. Here we have
assumed a temperature scaling relation as in Equation 2.11 of
\scite{kay99}. Assuming an underlying cosmology
($\Omega_0=0.3$, $\lambda_0=0$ in this case) and a value of the initial
power spectrum (here $n=-1.6$, in agreement with the measured APM galaxy power
spectrum, \pcite{tad98}), we can use the lack of observed evolution in the
L--T relation to constrain evolution in the cluster core entropy. 
This evolution of the entropy floor is represented by the
expected normalisation of the L--T relation changing, with redshift, as a
function of the value of $\epsilon$, as shown in
Fig.~\ref{fig:kayeps}. Here the degeneracy with $n$ is introduced in
the calculation of Equation 2.11 of \scite{kay99} and also with the effect of a changing epoch
of cluster formation as in Equations 2.12 and 2.13.
The lack of observed evolution (as demonstrated by the value of the $\eta$
parameter being consistent with zero), implies almost no evolution
of cluster core entropy to $z \sim 0.8$ given the assumed cosmology and
value of $n$. Table~\ref{tab:epsilon} details our calculated $\epsilon$
values in both an open and closed cosmology, for three different $n$
values. Also tabulated are the corresponding values required for pure
self-similar scaling in an $\Omega_0=1$ cosmology (\pcite{kai86}), given by
$\epsilon_{\rm ss}=-\frac{(n+7)}{(n+3)}$, which is strongly ruled out.

So far we have not attempted to place any cosmological constraints. To do this we 
need to break the degeneracy that allows evolution in the X-ray properties 
to be explained via either gravitational or gas-phase phenomena. This 
requires the introduction of a second dataset. The XLF, for instance, 
allows the values of $n$ and $\epsilon$ to be constrained in an almost 
orthogonal manner to the L--T evolution constraints. This is demonstrated 
in fig. 3 of \scite{kay99}. If the XLF is non-evolving to high 
redshift, as several recent studies have found (\pcite{bur97}; \pcite{jon98b}; 
\pcite{ros98}; \pcite{ebe00}), then important 
constraints can be placed on the underlying cosmology. Given that the value of
$\epsilon$ is more tightly constrained in this work than in
\scite{kay99}, any fluctuation spectrum of $n<-1.5$ requires a Universe with a
low value of $\Omega_0$ ($\Omega_0 < 0.5$). Correspondingly, values of $n
\sim -1$ are required to be consistent with the observed XLF and L--T relation
evolutionary constraints in an $\Omega_0=1$ universe. Allowing for the
possibility of some
negative evolution in the XLF at the highest luminosities, as some surveys find
(\pcite{hen92}, \pcite{vik98a}), may relax these constraints slightly,
although the evolution at these high luminosities is not observationally
well determined.

\begin{table}
\begin{minipage}[c]{8cm}
\center{\caption{\label{tab:epsilon}Constraints on the entropy evolution of cluster cores, as
  represented by the $\epsilon$ parameter. Values are listed for 3
  different values of the primordial fluctuation power spectrum, $n$, and in two different
  cosmologies. $\epsilon$ values predicted for pure self-similar scaling
  (Kaiser 1986) are also given.}} 
\begin{tabular}{|c|c|c|c|}\hline
 & $\epsilon \pm0.51 (1\sigma)$& $\epsilon \pm0.51 (1\sigma)$
& $\epsilon$ (self-similar) \\ 
$n$ value & $\Omega_0=0.3 , \lambda_0=0$ & $\Omega_0=1$  & $\Omega_0=1$ \\ \hline
-1 & -0.76 & 0.28 & -3.00 \\
-1.6 & -0.55 & 0.74 & -3.86 \\
-2 & -0.26 & 1.35 & -5.00 \\ \hline
\end{tabular}
\end{minipage}
\end{table}

\subsection{Cooling flows}

The abundance of cooling flows within the WARPS cluster survey could have a
large effect on the estimated cluster luminosity--temperature
relation. \scite{all98} showed that when the most X-ray luminous clusters
were modelled isothermally,
non-cooling flow clusters had significantly higher temperatures
than cooling flow clusters. Furthermore, they found a significant change
in the best-fitting slope of the cluster L--T relation when cooling flow
clusters were analysed with an isothermal model, as opposed to a model
with a central cooling flow component. They claimed a change in slope from
$\rm L \propto T^3$ to approximately $\rm L \propto T^2$ when
cooling flows were taken into account.

Due to the low signal-to-noise of the $ROSAT$ X-ray data used in this work,
fitting the data for a cooling flow component is not possible. However, we
were able to estimate the fraction of clusters in our sample that exhibit
evidence for a cooling flow. It has been shown that most clusters ($\ge 70$
 per cent) have or can form cooling flows (e.g. \pcite{edg92};
\pcite{per98}), though the cooling flow size is proportional to the X-ray
luminosity. Some cooling flows also exhibit other
characteristics, most notably optical line emission from the BCGs. The
fraction of clusters in X-ray 
flux limited surveys displaying these properties is around
33 per cent (e.g. \pcite{edg92}; \pcite{don92}), though this fraction is highly
dependent on the quality of the optical
spectra. Less than this level of cooling flow activity in the
WARPS sample would be a good indication that our survey does not
over-sample cooling flows.

The optical spectra obtained to measure galaxy redshifts allowed analysis of line emission in the BCG of each
cluster. In a preliminary analysis, H$\alpha$ or [OII]$\lambda3727$
emission lines were detected in 4 of the 26
BCG spectra analysed. This implies a fraction of 16 per cent cooling
flow clusters in our survey. This might be expected if the frequency of
cooling flows exhibiting optical emission lines increased in more X-ray
luminous systems. 

Cooling flows would produce a lower temperature estimate for the same
X-ray luminosity, thus they would, in effect, mimic evolution in the WARPS
results in the direction predicted for $\Omega_0=1$, assuming no core
entropy evolution (\pcite{bow97}). Removal of the effects of a fraction of the sample
containing cooling flows would hence favour lower values of $\Omega_0$,
strengthening our conclusion.

\section{Conclusions}

We have presented a cluster X-ray luminosity--temperature relation out to high
redshifts ($z \sim 0.8$), based on $ROSAT$ PSPC spectra of WARPS
clusters. Due to the low signal-to-noise of many of our $ROSAT$ PSPC
spectra, a multiple fitting technique was adopted in order to constrain
temperatures. The 91 clusters analysed were binned into redshift and
luminosity bins, and spectrally fit to a characteristic temperature.

In order to analyse the evolution of the L--T relation, a large sample of
low redshift ($z<0.2$) clusters was compiled from the literature. The best-fitting
power law to the data at low redshift and at T$>$1 keV was found to be a power law of index $\rm L
\propto T^{3.15\pm0.06}$ (90 per cent confidence). This relation was then compared with the high
redshift measurements to constrain possible
evolution in the normalisation of the L--T relation.

The binned $ROSAT$ measurements of WARPS clusters were consistent with no
evolution of the L--T relation to $z \sim 0.8$. When combined with previous cluster measurements from the
literature they agree with the no evolution result of \scite{don99}, but
give stronger constraints; $\eta=0.19\pm0.38$ for $\Omega_0=1$, or
$\eta=0.60\pm0.38$ for $\Omega_0=0.3$ where $\rm L_{Bol} \propto {(1+\it
  z)}^{\eta} \rm T^{3.15}$. Using the entropy driven evolution model of
\scite{bow97} and \scite{kay99}, a
limit to the implied entropy evolution of cluster cores was found.
Whilst the level of entropy evolution is slightly dependent on the value of
the primordial density fluctuation spectrum, $n$, low amounts of evolution were
favoured. The pure self-similar scaling expectations of
\scite{kai86} were found to be strongly excluded. 

The lack of evolution in the WARPS L--T relation, in conjunction with other
recent results, such as a non-evolving XLF, implies either
an open cosmology ($\Omega_0<1$) or one closed by a cosmological
constant, assuming $n<-1$. The forthcoming X-ray missions Chandra and XMM
will allow more accurate temperature estimates of clusters, out to high
redshifts, which will greatly improve our understanding of the L--T relation
and its cosmological implications.

\section{Acknowledgments}

The authors would like to thank Megan Donahue and collaborators for advance
access to their cluster measurements. Thanks to Scott Kay for help with the
entropy evolution model. We thank the Birmingham galaxies and
cluster group for useful discussions and input.
The data used in this work have been obtained from the Leicester database
and archive service (LEDAS). Computing facilities provided by the STARLINK
project have been used in this work. BWF acknowledges the receipt of a
PPARC studentship and LRJ also acknowledges the support of PPARC. HE
acknowledges financial support from SAO contract SV4-64008 and NASA LTSA
grant NAG 5-8253.

\bibliographystyle{mnras}
\bibliography{clusters}

\end{document}